\documentclass[%
 aip,
 jcp,%
 amsmath,amssymb,
reprint,%
author-numerical,%
floatfix
]{revtex4-1}

\usepackage{amssymb,amsfonts,amsmath,amsthm}
\usepackage{amstext}
\usepackage{bm}
\usepackage{lineno}
\usepackage{verbatim}
\usepackage{caption}
\usepackage{float}
\usepackage{graphicx}
\usepackage{wrapfig}
\usepackage{sidecap}
\usepackage{subfigure}
\usepackage{dcolumn}

\usepackage{color}
\usepackage{cleveref}
\usepackage{natbib}
\def\Hh{\mathcal{H}}

\def\E{\mathbb{E}}

\def\R{\mathbb{R}}

\def\d{d}

\def\dx{\d x}

\def\EXP#1{e^{#1}}

\def\EXPECT{{\mathbb{E}}}

\def\RELENT#1#2{\Rr\left(#1|#2\right)}

\def\WEAKLY{{\rightharpoonup}}

\def\COMMA{\,,}             
\def\PERIOD{\,.}            
\def\SEP{{\,|\,}}           

\def\VIZ#1{(\ref{#1})}      

\def\LAWEXACT{P}
\def\LAWAPPROX{Q}
\def\CONFNEW{\sigma'}

\def\PATHS{\LAWEXACT_{[0,T]}}
\def\PATHSAPP{\LAWAPPROX_{[0,T]}}
\def\PATHSAPPPAR{{{\LAWAPPROX}^\theta_{[0,T]}}}

\def\RELENT#1#2{\mathcal{R}\left({#1}\SEP{#2}\right)}
\def\ENTRATE#1#2{\mathcal{H}({#1}\SEP{#2})}

\def\BARIT#1{{\bar {#1}}}

\def\LATT{{\Lambda}_N}

\def\SIGMA{{\mathcal{S}_N}}
\def\STATESP{\Sigma}                

\def\COP{\mathbf{T}}                

\def\COPP{\mathbf{\Pi}}
\def\INC{\Delta}
\def\INCTH{\Delta^\theta}

\def\FISHERR{\mathbf{F_{\mathcal{H}}}}

\newtheorem{remark}{Remark}[section]

\begin{document}

\title{Information-theoretic tools for parametrized coarse-graining of non-equilibrium extended systems}
\thanks{Submitted to the Journal of Chemical Physics.}

\author{Markos A. Katsoulakis}
\email[Corresponding author:]{markos@math.umass.edu}
\affiliation{Department of Mathematics and Statistics, University of Massachusetts, Amherst, MA 01003 
}%

\author{Petr Plech\'a\v{c}}
\email{plechac@math.udel.edu}
\affiliation{Department of Mathematical Sciences, University of Delaware, Newark, DE 19716 
}%

\date{\today}

\begin{abstract}
In this paper we focus on the  development
of new methods suitable for efficient and reliable coarse-graining  of  {\it non-equilibrium}  molecular systems. 
In this context, 
we propose   error estimation and controlled-fidelity model reduction 
methods based on Path-Space Information Theory,   combined with statistical parametric estimation
of rates for non-equilibrium stationary processes. 
                  The approach we propose  extends the applicability 
                  of existing  information-based methods for deriving parametrized coarse-grained models to Non-Equilibrium systems with Stationary States (NESS). 
                  In the context of coarse-graining it allows for constructing optimal parametrized Markovian  coarse-grained dynamics within a parametric family, 
                  by minimizing information loss (due to coarse-graining) on the path space. 
Furthermore, we propose an asymptotically equivalent method--related to maximum likelihood estimators  for stochastic processes--where the  coarse-graining is obtained
by  optimizing the information 
content in  path space of the coarse variables,  with respect to the projected   computational data from a fine-scale simulation.
Finally, the associated  path-space  Fisher Information Matrix can provide confidence intervals  for the corresponding parameter estimators.   
We demonstrate the proposed  coarse-graining method in  (a) non-equilibrium systems with  diffusing  interacting particles, driven by out-of-equilibrium boundary conditions,
as well as (b) multi-scale diffusions and  their well-studied corresponding stochastic averaging limits, comparing them to our proposed methodologies.           
 
\end{abstract}


%


\maketitle

%
%
\section{Introduction}\label{intro}

 Non-equilibrium  systems at transient or steady state regimes are 
typical in applied science and  engineering, and are the result of coupling between different physicochemical  mechanisms, driven by external couplings or boundary conditions.
Typical examples include  reaction-diffusion systems in heteroepitaxial catalytic materials, polymeric flows  and separation processes in microporous materials,
\cite{SV2012, Briels_Rev:11, zeolites_rev}.
In this paper we  develop reliable model-reduction methods, i.e., having controlled fidelity of approximation, and  capable to handle extended, {\it non-equilibrium} statistical mechanics models. 
These  coarse-graining methods allow for constructing optimal parametrized Markovian  coarse-grained dynamics within a parametric family, 
                  by minimizing information loss (due to coarse-graining) on the path space. 
Model-reduction (or coarse-graining)  approaches can be often described in the context of parameter estimation in parametrized statistical models. 
However, atomistic models of materials lead to high-dimensional probability distributions and/or stochastic processes to which the standard methods of 
statistical inference and  model discrimination are not directly applicable.
The emphasis on information theory tools is also partly justified since often we are interested in   probability 
density functions (PDF), typically non-Gaussian, 
due to the significance of tail events in complex systems. A primary focus of this paper is on systems with Non-Equilibrium
Steady States (NESS), i.e., systems in which a steady state is reached but the detailed balance condition is violated and  
explicit formulas for the stationary distribution, e.g., in the form of a Gibbs distribution, are not available. 
    
Information-theoretic methods for the  analysis of stochastic models typically employ  entropy-based tools for analyzing 
and estimating a distance between (probability) measures.  
In particular, the relative entropy (Kullback-Leibler divergence) of two probability measures $\mu(dx) = \mu(x)\,dx$ and $\nu(dx) = \nu(x)\,dx$
$$
\RELENT{\mu}{\nu} = \int \mu(x) \log\frac{\mu(x)}{\nu(x)}\,dx
$$
allows us to define a pseudo-distance between two measures. 
A key property of the   relative entropy $\RELENT{P}{Q}$ 
is that $\RELENT{P}{Q} \ge 0$ with equality if and only if $P=Q$, 
which  allows us to view relative entropy  as a ``distance" (more precisely a semi-metric)  
between two probability measures $P$ and $Q$. Moreover, from an information theory
perspective \cite{Cover},  the relative entropy measures {\it loss/change of information}.
Relative entropy for high-dimensional systems was used as measure
of loss of information in coarse-graining \cite{Kats:Trashorras:06, KPRT1, Arnst:08}, 
and sensitivity analysis for climate modeling problems \cite{Majda:10}. 

Using entropy-based analytical tools has
proved essential for deriving rigorous results for passage from interacting particle
models to mean-field descriptions, \cite{Landim}.
The application of relative entropy methods to the error analysis of coarse-graining of stochastic particle
systems have been introduced and studied in  \cite{KPRT1,KPRT2,KPRT3,KMV,KV}. Aside  of this rigorous numerical analysis  direction, 
entropy-based computational techniques were also developed  and used for constructing approximations of coarse-grained (effective) potentials for models
of large biomolecules and polymeric systems (fluids, melts). Optimal parametrization of effective potentials based on
minimizing the relative entropy between {\em equilibrium} Gibbs states, e.g.,\cite{shell1, shell2, zabaras}, extended previously developed inverse Monte Carlo 
methods, primarily based on force matching approaches, used in coarse-graining of macromolecules (see, e.g., \cite{kremer,mp}). 
In \cite{Espanol:11} an extension to dynamics is proposed in the context of  Fokker-Planck  equations, by considering 
the  corresponding relative entropy for discrete-time approximations  of the transition probabilities. 
Furthermore, relative entropy was used as means to 
improve model fidelity in a parametric, multi-model  approximation framework of complex  dynamical systems, at least  when the model's steady-state distributions are explicitly known, 
 \cite{Majda:11}. Overall, such parametrization techniques are focusing on 
                  systems with  a known steady state,  such as  a Gibbs   equilibrium distribution.
More specifically, computational implementations of 
optimal parametrization in the inverse Monte Carlo methods is relatively straightforward for equilibrium systems in which the
best-fit procedure is applied to an {\em explicitly known} equilibrium distribution   and  where relative entropy is explicitly  computable.

On the other hand, this is not the case in  non-equilibrium systems,
even at a steady state where typically we do not have a Gibbs structure and the steady-state distribution  is unknown altogether, setting up one of the primary challenges for this paper.
Indeed, here  we show that, in non-equilibrium systems  the general  information theory  ideas based on Kullback-Leibler divergence are still applicable 
but they have to be properly formulated 
in the context of  Non-equilibrium Statistical Mechanics by focusing on the  probability distribution of the entire time series, i.e.,   on the {\em path space} of the underlying 
stochastic processes.
We show that, surprisingly, such a path-space relative entropy formulation  is:  (a) general in the sense that it applies to any Markovian models (e.g. Langevin dynamics, Kinetic Monte Carlo, etc),  and (b) is easily computable as an ergodic average  in terms of the {\em Relative Entropy Rate\,} (RER),  therefore allowing us to 
construct  optimal parametrized Markovian 
 coarse-grained dynamics for large classes of models.  This procedure involves the minimization  of information loss in path space, where information is inadvertently lost due to coarse-graining procedure. 
 In fact, the proposed parametrization scheme in \cite{Espanol:11} is mathematically  justified by reformulating it on the path space 
 using the Relative Entropy Rate and it is  a specific, but reversible (i.e., it has a Gibbs steady state) example of our methodology.
 Furthermore, we propose an asymptotically equivalent method to RER minimization, which is related to  to maximum likelihood estimators  for stochastic processes,
 and where the optimal  coarse-graining is obtained    by  optimizing the information 
content in  path space of the coarse variables,  with respect to the projected   computational data from a fine-scale simulation.
 Finally,  the   path-space  Fisher Information Matrix (FIM) derived from Relative Entropy Rate (RER)  can   provide  confidence intervals  for the corresponding  statistical  estimators of the optimal parameter
obtained through the minimization problem.

 {  The Relative Entropy Rate (RER) was  earlier  proposed as a (pseudo) metric in order to evaluate the convergence of adaptive sampling 
 schemes for Markov State Models (MSM)\cite{Bowman:2010}; the proposed coarse-graining  perspective, which is related to Maximum Likelihood Estimators (MLE), e.g.,  \VIZ{Path-L}, could provide new insights in comparing MSMs since it only requires fine-scale data--and not explicitly a fine-scale  MSM for estimating the RER.  Furthermore, RER and in particular the  path-space FIM where introduced recently as  gradient-free sensitivity analysis tools for complex, non-equilibrium stochastic systems with a wide range of applicability to reaction networks, lattice Kinetic Monte Carlo and Langevin dynamics \cite{PK2013}.  
 Unlike   the coarse-graining setting we propose  here and where models are set in different state spaces depending on their granularity,  in the latter work the same state space is  assumed between 
 compared models.}

The paper is structured as follows. In Section~\ref{path_space} we formulate the path-space information theory tools used in the paper,  including the key concept of Relative Entropy Rate.
In Section~\ref{parametrization} we present the parameterization method of coarse-grained dynamics and the connections with maximum likelihood estimators and the Fisher Information Matrix.
In Section~\ref{stat_est} we briefly discuss statistical  estimators for RER and FIM. Finally, in Section~\ref{benchmark}
we demonstrate the proposed  coarse-graining method in     (a) non-equilibrium systems with  diffusing  interacting particles, driven by out-of-equilibrium boundary conditions,
as well as (b) multi-scale diffusions and  their well-studied corresponding stochastic averaging limits, comparing them to our proposed methodologies.

%
%
\section{Relative Entropy Rate, path space information theory   and error quantification for non-equilibrium systems}\label{path_space}



First,  we formulate a general entropy-based error analysis for coarse-graining, dimensional reduction and parametrization of high-dimensional Markov processes,
simulated by Kinetic Monte Carlo (KMC) and Langevin Dynamics. 
Typically such systems have  Non-Equilibrium Steady States (NESS)  for which {\em detailed balance fails} as they are {\em irreversible}.
The stationary distributions are not known explicitly and have to be studied computationally.
{ 
\subsection{ Coarse-grained   models}
We consider a parameterized class of coarse-grained   
Markov processes $\{\eta_t\}_{t \ge 0}$, 
associated with the fine scale   stochastic
process $\{\sigma_t\}_{t \ge 0}$. The coarse-graining procedure  is based on projecting the
microscopic space $\Sigma$ into a coarse space $\BARIT\Sigma$ with less degrees
of freedom. We denote the
coarse space variables 
\begin{equation}\label{cgmap}
\eta=\COP\sigma\, , \quad\mbox{where} \quad \COP: \Sigma \to \bar{\Sigma}
\end{equation}
 is a coarse-graining (projection) operator, see also \VIZ{cgmap_latt} for a specific example.
%
In general, given a fine-scale probability measure 
 ${\mu}_{N}(d\sigma)$ defined on the fine-grained configuration space $\Sigma$, 
 then we can define the exact (renormalized) coarse-grained probability measure  $\bar \mu_{M}^{\rm app}(d \eta)$ on the coarse space $\bar \Sigma$:
 \begin{equation}\label{exact_CG}
\bar{\mu}_{M}(d\eta)= \int_{\{\sigma:\COP\sigma=\eta \}} {\mu}_{N}(d\sigma)\, .
\end{equation}
However, the exact CG probability cannot be explicitly calculated , except in trivial cases. Typically,  we consider classes of parametric  CG models which approximate the exact CG model, see for instance the quantification of such approximations using relative entropy \cite{KPRT1}.

In the case of continuous time processes such as Kinetic Monte Carlo, the coarse-grained stochastic 
process is defined in terms of coarse transition  rates  $ \BARIT{c}(\eta,\eta')$
which captures macroscopic information from the fine scale rates $c(\sigma,\sigma')$. 
For example, for stochastic lattice systems, approximate coarse rate 
functions are explicitly known from coarse graining (CG) techniques  of \cite{KMV, KV, sinno}, see \VIZ{cgrates}. 
Similarly, when we consider temporally discretized stochastic processes such as Langevin Dynamics, the coarse-grained process is 
given in terms of transition probabilities
$ \BARIT{p}(\eta,\eta')$
which capture macroscopic information from the fine scale transition probabilities  $p(\sigma,\sigma') $.

\noindent
{\it Interpolated dynamics.} Given coarse-grained dynamics we can always construct corresponding microscopic dynamics.
For example,
given  coarse-grained  transition probabilities $ \BARIT{p}(\eta,\eta')$ with corresponding stationary distribution $\BARIT{\mu}$, 
where the latter is typically unknown in non-equilibrium systems, we define the corresponding fine-scale rates
\begin{equation}\label{recon_uniform}
 q(\sigma,\sigma'):= U(\sigma'|\COP\sigma')\BARIT{p}(\COP\sigma,\COP\sigma')\COMMA
 \end{equation}
 where $U(\sigma'|\eta')={1 \over |\{\sigma:\COP\sigma=\eta'\}|}$ ($|\cdot |$ denotes the cardinality)  is the uniform conditional distribution over all fine-scale states $\sigma$ corresponding to the same coarse-grained state $\eta'$.
 Clearly \VIZ{recon_uniform} is properly normalized, while a more general formulation can be found in Appendix~\ref{appdx}, \VIZ{recon}.
In \VIZ{recon_uniform} we apply a piece-wise constant interpolation for all microscopic states $\sigma$ (reps. $\sigma'$) corresponding to the same 
coarse state $\eta$ (reps. $\eta'$) and thus transitions to these states occur with the same probability rates.
 The reconstruction step \VIZ{recon_uniform} is necessary when we want to compare fine and coarse processes on the path space   
 in terms of the relative entropy rate,
 since both processes need to be defined on the same probability space, see for example \VIZ{entrate-optim} below. 
 In this paper, for the sake of simplicity, we assume that all reconstructions are based on \VIZ{recon_uniform}. 
 The reconstruction is obviously not unique and we refer to Appendix~\ref{appdx} for the mathematical details, see also \cite{TT}. 
  

\subsection{Relative Entropy and Error Quantification in Non-equilibrium Systems}
Although all  systems we consider here are {\em ergodic}, i.e., they have a unique steady state distribution, we typically assume they  only have a    Non-Equilibrium Steady State (NESS) due to a lack of  detailed balance\cite{BLR}. Inevitably, in such systems 
the stationary distribution is  not known explicitly and can be studied  primarily  computationally, at least in systems which are not small perturbations from equilibrium.
}
Quantifying and controlling the
coarse-graining error in such high-dimensional stochastic  systems  can be achieved by   developing  {\em computable and efficient}   methods for estimating distances of probability 
measures on the path space.
The relative entropy between two path measures $\PATHS$ and $\PATHSAPP$ (see \VIZ{path-measure} for a specific example)
for the processes on the interval $[0,T]$ is 
\begin{equation}\label{path-relentropy}
\RELENT{\PATHS}{\PATHSAPP} = \EXPECT_{\PATHS}\left[\log\frac{d\PATHS}{d\PATHSAPP}\right]\COMMA
\end{equation}
where $\tfrac{d\PATHS}{d\PATHSAPP}$ is the Radon-Nikodym derivative of $\PATHS$ with respect to $\PATHSAPP$. 
If these
probability measures have probability densities $p$,
$q$ respectively, \VIZ{path-relentropy} becomes 
 $\RELENT{{\PATHS}}{{\PATHSAPP}}= \int p\log\left(\tfrac{p}{q}\right)$.
In the setting of coarse-graining or model-reduction the measure $\PATHS$ is associated with the exact process
and $\PATHSAPP$ with the approximating (coarse-grained) process. 

From an information theory perspective, the relative entropy measures the loss of information as we approximate the exact stochastic process $\PATHS$
with the coarse-grained one $\PATHSAPP$.
 In general the relative entropy \VIZ{path-relentropy} in this dynamic
setting is not a computable object; we refer for instance to related formulas in the Shannon-MacMillan-Breiman Theorem, \cite{Cover}. 
However,  as we show next, in practically relevant cases of {\it stationary} Markov processes 
we can work with the {\em relative entropy rate}
\begin{equation}\label{rate-relentropy}
  \ENTRATE{P}{Q} = \lim_{T\to \infty} \frac{1}{T} \RELENT{\PATHS}{\PATHSAPP} \COMMA
\end{equation}
where $P$ and $Q$ denote the distributions of the corresponding stationary processes.

  
\noindent
{\it Relative Entropy Rate for Markov Chains.} In order to explain the basic concept we restrict to the case of two Markov chains, $\{\sigma_n\}_{n\geq 0}$, $\{\tilde \sigma_n\}_{n\geq 0}$
on the countable state space $\STATESP$, defined
by the transition probability kernels $p(\sigma,\sigma')$ and $q(\sigma,\sigma')$.  
A  typical example would be the embedded Markov chain used for KMC simulations of a continuous time Markov chain. Similarly, in the case of a continuous state space,
a temporal discretization of a Langevin process, 
leads to a Markov process with the transition kernel $p(\sigma,d\sigma') = p_{\Delta t}(\sigma,\sigma')\,dx'$ defined
by the time-discretization scheme of the underlying stochastic dynamics.
We assume that the
initial states are from the invariant distributions $\mu(\sigma)$ and $\nu(\sigma)$. The path measure
defining the probability of a path $(\sigma_0, \sigma_1, \dots, \sigma_T)$ is then  
\begin{equation}\label{path-measure}
P(\sigma_0,\dots,\sigma_T) = \mu(\sigma_0) p(\sigma_0,\sigma_1)\dots p(\sigma_{T-1},\sigma_T)\COMMA
\end{equation}
and similarly for the measure $Q(\sigma_0,\dots,\sigma_T)$. The Radon-Nikodym derivative is easily computed
$$
\frac{dP}{dQ} = 
\frac{\mu(\sigma_0)\prod_{i=0}^{T-1} p(\sigma_i,\sigma_{i+1})}%
     {\nu(\sigma_0)\prod_{i=0}^{T-1} q(\sigma_i,\sigma_{i+1})}\PERIOD
$$
Using the fact that the processes are stationary with invariant measures $\mu$ and $\nu$, 
we obtain an expression for the relative entropy
\begin{equation}\label{pathrelentropy}
\RELENT{P}{Q}  = T  \,\EXPECT_{\mu} \left[ \sum_{\sigma'\in\STATESP} p(\sigma,\sigma') \log \frac{p(\sigma,\sigma')}{q(\sigma,\sigma')}\right] + \RELENT{\mu}{\nu}\COMMA
\end{equation}
and thus the relative entropy rate is given explicitly as
\begin{equation}\label{entrate}
\ENTRATE{P}{Q} = \sum_{\sigma\in\STATESP} \mu(\sigma) \sum_{\sigma'\in\STATESP} p(\sigma,\sigma') \log \frac{p(\sigma,\sigma')}{q(\sigma,\sigma')}]\PERIOD
\end{equation}
We will refer from now on to the quantity \VIZ{entrate} as the {\em Relative Entropy Rate}
(RER), which can be thought as the change in information per unit time. Notice that RER
has the correct time scaling since it is actually independent of the interval $[0,T]$. Furthermore, it has the following key features that make it
a crucial observable for simulating and coarse-graining complex dynamics:

\begin{description}
 \item[{\rm (i)}] The RER formula 
   \VIZ{entrate} provides a computable observable that can be sampled from the steady
   state $\mu$ in terms of conventional  Kinetic Monte Carlo (KMC),  bypassing  the need for a histogram
   or an explicit formula for the high-dimensional probabilities involved in \VIZ{path-relentropy}.
 \item[{\rm (ii)}] In stationary  regimes,  when $T\gg 1$  in \VIZ{pathrelentropy},
   the term $\RELENT{\mu}{\nu}$ becomes unimportant. This  is
   especially convenient since  $\mu$ and $\nu$ are typically not known
   explicitly in non-reversible systems, for instance  in   reaction-diffusion or driven-diffusion  KMC
   or non-reversible Langevin dynamics.
\end{description}
In view of these features, we readily see that if we consider   a Markov  chain $\{\tilde \sigma_n\}_{n\geq 0}$ as an approximation, e.g., a  coarse-graining, 
of the chain $\{\sigma_n\}_{n\geq 0}$, we can
estimate the loss of information at long times by computing $\ENTRATE{P}{\tilde P}$ as an ergodic average. 
This observation is the starting point of the proposed methodology and relies on the fact that 
the observable $\ENTRATE{P}{Q}$ is {\em computable};  efficient statistical estimators for \VIZ{entrate} are discussed
in Section~\ref{stat_est}. A similar calculation can be carried out for continuous time Markov Chains, as we see next.

\noindent
{\it Continuous Time Markov Chains and Kinetic Monte Carlo.}
In models of catalytic reactions or epitaxial growth the systems are often described by continuous time Markov chains (CTMC) that are simulated by KMC algorithms.
For example, the  microscopic Markov process $\{\sigma_t\}_{t\geq 0}$ describes the evolution of molecules on a substrate lattice. 
Mathematically 
the  continuous time 
Markov chain is defined  completely by specifying the local  transition rates $c^\theta(\sigma, \sigma')$ where 
$\theta\in\R^k$ is a vector of the model parameters.  The transition rates  determine  the  updates 
from any current state (configuration)  $\sigma_t=\sigma$ to a (random) new state $\sigma'$. 
In the context of the spatial models considered here, 
the  transition rates take the form $c^\theta(\sigma, \sigma')=c^\theta(x,\omega,\sigma)$, denoting by $x \in \LATT$ a lattice site
on a $d$-dimensional lattice $\LATT$ and 
$\omega\in\SIGMA_x$, where  $\SIGMA_x$ is the set of all possible configurations that correspond to an update in a neighborhood
of the site $x$. 
From local transition rates one defines the total rate $\lambda^\theta(\sigma)=\sum_{x \in \LATT}\sum_{\omega \in \SIGMA_x } 
c^\theta(x, \omega, \sigma)$, which is the intensity of the exponential waiting time for a jump from the state $\sigma$. 
The transition probabilities for the embedded Markov chain $\{S_n\}_{n\geq 0}$ are 
$p(\sigma, \CONFNEW;\theta)=\frac{c(x, \omega, \sigma;\theta)}{\lambda(\sigma;\theta)}$.
In other words once the exponential ``clock'' signals a jump, the system transitions from the state $\sigma$ 
to a new configuration $\CONFNEW$ with the probability $p(\sigma, \CONFNEW)$.
In the context of coarse-graining  we are led to finding an optimal parametrization for the rates $\tilde c(\sigma,\sigma';\theta)$
of a processes that approximates the dynamics given by the microscopic process $c(\sigma,\sigma')$. 
A similar calculation as in the case of Markov chains gives the analogue of the formula \VIZ{entrate}
\begin{equation}\label{entrate2}
\ENTRATE{P}{Q}= \E_\mu \left[\lambda (\sigma) -\tilde \lambda(\sigma;\theta) -
                                                 \sum_{\sigma'} c(\sigma, \sigma')\log \frac{
                                                 c(\sigma, \sigma')}{\tilde c(\sigma, \sigma';\theta)}\right]\COMMA
\end{equation}
where $\mu$ is the stationary distribution of the microscopic process and $\lambda$ denotes the total transition rate.
In \cite{KKPV} we used this quantity 
in order to quantify error in a two-level coarse-grained kinetic Monte Carlo method. Based on these considerations,  
we show in Section~\ref{parametrization}  that minimizing the error 
measured by \VIZ{entrate2} leads to a CTMC coarse-grained dynamics that best approximates long-time behavior of 
the microscopic process projected  to the coarse degrees of freedom. 

\begin{remark}
{\rm
We consider the special case where  the transition probability function of the Markov chain is sampled  directly from the invariant measure, i.e., 
$$
p(\sigma,\sigma') = \mu(\sigma'), \;\mbox{and $q(\sigma,\sigma') = \nu(\sigma')$, for all $\sigma,\sigma'\in\Sigma$.}
$$
This sampling   is equivalent to the fact that the path space samples in \VIZ{pathrelentropy}
are
independent and  identically distributed from the stationary probability distributions.  Then the RER between the path probabilities becomes the usual relative entropy between the
stationary distributions:
\begin{equation}
\ENTRATE{P}{Q}=\RELENT{\mu}{\nu}\, .
\end{equation}
Estimating RER using \VIZ{entrate} is far simpler than directly estimating the relative entropy $\RELENT{\mu}{\nu}$, since \VIZ{entrate} only involves local dynamics rather than 
the full steady state measure, which typically may not be available. Furthermore even when it is available in the form of a Gibbs state 
it will require computations that will typically  involve a full Hamiltonian,
\cite{shell1, KPRT2}.
}
\end{remark}

{ 
\begin{remark} We also note that the RER can be written as a relative entropy, inheriting all its properties \cite{Cover}, e.g., non negativity, convexity, etc. In fact, we can rewrite \VIZ{entrate} as
\begin{equation}\label{RER_RE}
\ENTRATE{P}{Q}=\RELENT{\mu\otimes p}{\mu\otimes q}\, ,
\end{equation}
where we define the product probability measures  as $\mu\otimes p(A \times B)=\sum_{\sigma \in A}\mu(\sigma)\sum_{\sigma' \in B}p(\sigma, \sigma')$. 
\end{remark}
}

\noindent
{\it Estimation and error  of observables.}
The estimates  on relative entropy and RER  can provide an upper bound for a large family of observable
functions through the Pinsker (or Csiszar-Kullback-Pinsker) inequality. The Pinsker inequality
states that the total variation norm between $\PATHS$ and $\PATHSAPP$
is bounded in terms  of the relative entropy, \cite{Cover}.
The Pinsker inequality  gives an estimate for a difference of the mean computed with respect to the distribution $P$ and $Q$
\begin{equation}
|\mathbb E_{\PATHS}[f] - \mathbb E_{\PATHSAPP}[f]| \leq ||f||_\infty \sqrt{2 \RELENT{\PATHS}{\PATHSAPP}}\, ,
\label{Pinsker:ineq}
\end{equation}
where $ ||f||_\infty=\max |f|$.
An important conclusion that is immediately
drawn from the above inequality is that if the relative entropy of a distribution with respect to another
distribution is small then the error between 
any bounded observable functions is also accordingly small. Using \VIZ{pathrelentropy} we readily obtain the estimate
\begin{equation}
\begin{aligned}
|\mathbb E_{\PATHS}[f] - & \mathbb E_{\PATHSAPP}[f]| \leq  \\
         &||f||_\infty \sqrt{2T}\sqrt{\ENTRATE{P}{Q}+{1 \over T}\RELENT{\mu}{\nu}}\, ,
\label{Pinsker:ineq2}
\end{aligned}
\end{equation}
involving the relative entropy rate \VIZ{entrate} or \VIZ{entrate2}. As in virtually all numerical analysis estimates for stochastic dynamical systems, 
the bound \VIZ{Pinsker:ineq2}
may not be sharp, but it is indicative of the error in the observables when the distribution $Q$ approximates $P$.
\section{Parametrization  of coarse-grained dynamics  and Inverse Dynamic Monte Carlo}\label{parametrization}

\subsection{Inverse Dynamic Monte Carlo methods.}
In many applications the coarse-grained models are defined by effective potentials or effective rates which are sought
in a family of parameter-dependent functions, \cite{kremer,mp, g219}. 
The parameters are then fitted by minimizing certain functionals that
attempt to capture different aspects of modeling errors, e.g., radial distribution functions in \cite{mp}.
Compared to such  Inverse Monte Carlo methods applied to equilibrium systems we cannot work directly with
equilibrium distributions since 
the  NESS  is not explicitly known. Thus
we apply the information-theoretic framework on the path space, i.e.,
on the approximating
measure $\PATHSAPP\equiv \PATHSAPPPAR$ that  depends on the parameters $\theta\in\R^k$, and which  are subsequently fitted using entropy
based criteria for the best approximation. 

The optimal parametrized coarse-grained transition probabilities  $q^{\theta^*}(\sigma, \sigma')$ are constructed  as follows.
First, given the parametrized coarse-grained  transition probabilities $ \BARIT{p}^\theta(\eta,\eta')$ we define the fine-scale 
projected rates $q^\theta(\sigma, \sigma')$, which can be defined, for instance, by \VIZ{recon_uniform} as 
\begin{equation}\label{recon_uniform2}
q^\theta(\sigma, \sigma')=U(\sigma'|\COP\sigma')\BARIT{p}^\theta(\COP\sigma,\COP\sigma')\, ,
\end{equation}
and  the corresponding coarse-grained path-distribution is
\begin{equation}\label{path-measure-CG}
Q^\theta(\sigma_0,\dots,\sigma_T) = \BARIT{\mu}(\COP\sigma_0) q^\theta(\sigma_0, \sigma_1)\dots q^\theta(\sigma_{T-1}, \sigma_T)\, .
\end{equation}
Subsequently the best-fit
can be obtained by minimizing the relative entropy rate, i.e., finding a solution
\begin{equation}\label{optim-problem}
   \theta^* = \mathrm{arg}\min_\theta \ENTRATE{P}{Q^\theta}\, ,
\end{equation}
where now we have that the RER is
\begin{equation}\label{entrate-optim}
\ENTRATE{P}{Q^\theta} = \sum_{\sigma\in\STATESP} \mu(\sigma) \sum_{\sigma'\in\STATESP} p(\sigma, \sigma') \log \frac{p(\sigma, \sigma')}{q^\theta(\sigma, \sigma')}]\PERIOD
\end{equation}
This optimization problem on one hand is similar to more common parametric inference in which the log-likelihood function is maximized,
and this perspective will be further clarified in Section~\ref{data}. Furthermore, due to the parametric identification of 
the coarse-grained dynamics, i.e., transition probabilities, or rates in the case of \VIZ{entrate2},  we refer to the proposed methodology  as
an  {\em Inverse Dynamic Monte Carlo} method in analogy to the Inverse Monte Carlo methods for equilibrium systems, \cite{kremer,mp, g219}.

The optimization algorithm for \VIZ{optim-problem} is based on iterative procedures that locate a solution $\theta^*$ of the
optimality condition $\nabla_\theta  \ENTRATE{P}{Q^\theta} = 0$ 
\begin{equation}\label{iteration}
   \theta^{(n+1)} = \theta^{(n)} - \frac{\alpha}{n} G^{(n+1)}\COMMA
\end{equation}
for some $\alpha >0$ and $G^{(n+1)}$ being a suitable approximation of the gradient
$\nabla_\theta \ENTRATE{P}{Q^\theta}$, more precisely 
$\EXPECT[G^{(n+1)}\SEP G^{(0)}, \theta^{(0)},\dots, G^{(n)}, \theta^{(n)}] = \nabla_\theta\ENTRATE{P}{Q^\theta}$. 
The crucial ingredient  of this algorithm is an efficient and reliable estimator for the sequence
$G^{(n)}$ of the gradient estimates. Similar to the deterministic case the minimization can be accelerated by combining 
this step with the Newton-Raphson method and choosing the vector $G$ as 
\begin{equation}\label{NR-FIM}
G^n = \mathrm{Hess}(\ENTRATE{P}{Q^{\theta^n}})^{-1}\nabla_\theta \ENTRATE{P}{Q^{\theta^n}}\PERIOD
\end{equation}
While the evaluation of the Hessian $\mathrm{Hess}(\ENTRATE{P}{Q^{\theta^n}})$ presents an additional
computational cost,  it also offers additional information about the parametrization, sensitivity and
{\em identifiability} of the approximating model, \cite{PK2013}. Indeed the first and the second derivatives of the rate function $\ENTRATE{P}{Q^{\theta^n}}$
are of the form: 
\begin{equation}
\label{Gradient:MC}
\nabla_\theta(\ENTRATE{P}{Q^{\theta}})=
-\mathbb E_{\mu} \left[\sum_{\sigma'} p(\sigma,\sigma) \nabla_\theta
\log q^\theta(\sigma,\sigma') \right]\COMMA
\end{equation}
and $\FISHERR\big({{Q^{\theta}}}\big)=\mathrm{Hess}(\ENTRATE{P}{Q^{\theta}})$, where
\begin{equation}
\label{FIM:MC}
\FISHERR\big({{Q^{\theta}}}\big) = 
-\mathbb E_{\mu} \left[\sum_{\sigma'} p(\sigma,\sigma) \nabla^2_\theta
 \log q^\theta(\sigma,\sigma')  \right]\PERIOD
\end{equation}
The Hessian can be interpreted as a dynamic analogue of the Fisher Information Matrix (FIM) $\FISHERR\big({{Q^{\theta}}}\big)$
on the path space. A similar quantity, in the context of sensitivity analysis, 
was recently considered in \cite{PK2013}, where 
%
the authors also  developed efficient statistical estimators for the derivatives of RER
$\partial_{\theta_k} \ENTRATE{P}{Q^\theta}$ and $\partial^2_{\theta_i\theta_j} \ENTRATE{P}{Q^\theta}$. 
We discuss related   estimators in Section~\ref{stat_est}.

\begin{remark}
{\rm
%
The proposed approach carries sufficient level of generality in order to be applicable to a wide class of stochastic  processes, e.g., Langevin dynamics and  KMC,
without restriction to the dimension of the system, provided  scalable efficient simulators are available  to simulate the  observables, \VIZ{entrate} and \VIZ{entrate2}.
The proposed parametrized coarse-graining is applicable to any system for which a {\em  parametrized coarse-grained  models} are available, 
e.g., in coarse-graining of macromolecules and biomembranes, \cite{Briels_Rev:11, kremer, mp}.
An obvious obstacle is that the path measure $\PATHS$ is absolutely continuous with respect to $\PATHSAPP$, however,
it does not significantly restrict the class of relevant applications as we typically deal with KMC or Markov Chain approximations resulting from 
a discretization of Molecular Dynamics with noise.
In the latter case,  Markov chains obtained
by numerical approximations of stochastic differential equations (SDEs) allow us to compute RER through \VIZ{entrate} and  can be used for quantification
of errors or inverse Monte Carlo fitting for non-equilibrium or irreversible models in Section~\ref{parametrization}.

{ 
For example, the overdamped molecular dynamics with positions $x\in\R^d$ and the forcefield $a(x)\in\R^d$ 
at the inverse temperature $\beta>0$ is
a diffusion process $X_t$ given by the stochastic differential equations driven by the $d$-dimensional Wiener process $W_t$ 
 $$
    dX_t =  a(X_t) \,dt + \sqrt{2\beta^{-1}} dW_t\COMMA\;\;\; X_0 = x \PERIOD
 $$
 We assume that the drift field $a(x)$ satisfies standard conditions that guarantee existence of solutions for all $X_0 = x$ and 
 the process is ergodic with the stationary distribution $\mu(x)\,dx$.
 The stochastic differential equation can be  discretized by the Euler scheme with the time-step $h$
 \begin{equation}\label{Euler}
 X^{n+1} =  X^n + a(X^n) h + \sqrt{2\beta^{-1}} Z \sqrt{h}\COMMA
 \end{equation}
 where $Z \sim N(0,1)$ is a random increment from the standard normal distribution. 
 The Euler scheme discretization defines the Markov chain $X^n$ with the transition kernel 
 $$
 p_h(x,x') dx' \sim \EXP{-\frac{\beta}{h} |x' - x - h a(x)|^2}\dx'\PERIOD
 $$
The time-continuous case presents technical difficulties that we do not address here, instead we demonstrate application of the
proposed method at the level of the approximating Markov chain only. 
The case where the process is driven by a multiplicative noise $\sigma(X_t) dW_t$ can be handled in a similar way using a discrete scheme. 
For the sake of simplicity we define the coarse-graining operator $\COPP$ as an orthogonal projection from the state space $\R^d$ to a subspace $\R^m$, 
and we write $x = \COPP x + \COPP^\perp x$, denoting $\bar x\equiv \COPP x\in\R^m$, $\tilde x\equiv \COPP^\perp x\in\R^{d-m}$.
 The reduced model is then viewed as an approximation of the projected Markov chain
 \begin{equation}\label{PROJEuler}
    \COPP X^{n+1} =  \COPP X^n + \COPP a(X^n) h + \sqrt{2\beta^{-1}}\COPP Z  \sqrt{h}\COMMA
 \end{equation}
 by
 \begin{equation}\label{CGEuler}
    \bar X^{n+1} =  \bar X^n + \bar a(\bar X^n;\theta) h + \sqrt{2\beta^{-1}} \bar Z \sqrt{h}\COMMA
 \end{equation}
 where the Gaussian increments are $\bar Z \sim  N(0,\COPP\COPP^T)$.
 Denoting
 $\INC(x)\equiv x + a(x) h$ and $\INCTH(\bar x)\equiv \bar x + \bar a(\bar x;\theta) h$ the drift increments in \VIZ{Euler}
 and \VIZ{CGEuler} respectively we have the transition kernel of $\bar X^n$ given by 
 $\bar p_h(\bar x,\bar x';\theta) \sim \EXP{-\frac{\beta}{h} |\bar x' - \INCTH(\bar x)|^2}\, d\bar x'$. 
 Given the transition kernel $\bar p_h$ of the coarse-grained chain $\bar X^n$ we define, similarly as in \VIZ{recon_uniform2},
 the transition kernel of a reconstructed
 chain on the original state space $\R^d$
 $$
 q_h(x,x';\theta) = \bar p_h(\COPP x,\COPP x';\theta) \nu(x'|\COPP x')\PERIOD
 $$
 As long as the reconstruction measure $\nu$ does not depend on the parameters $\theta$ the particular
 choice of $\nu$ does not enter the optimality condition for $\ENTRATE{P}{P^\theta}$.
 Hence the relative entropy rate is given by
  \begin{equation}
    \ENTRATE{P}{P^\theta} = \int\int \mu(x) p_h(x,x') \log{\frac{p_h(x,x')}{q_h(x,x';\theta)}}\dx'\dx\PERIOD
  \end{equation}
 Note that due to the choice of the orthogonal projection $\COPP$ the transition kernel for \VIZ{Euler} of the full model becomes
 $$
 p_h(x,x') = \frac{1}{\bar Z} \EXP{-\frac{\beta}{h} |\COPP x' - \COPP\INC(x)|^2}\, d\bar x'\times
             \frac{1}{\tilde Z} \EXP{-\frac{\beta}{h} |\COPP^\perp x' - \COPP^\perp\INC(x)|^2}\, d\tilde x'\COMMA
 $$
 and this factorization into a product simplifies the evaluation of the necessary condition
 for a minimizer of $\min_\theta \ENTRATE{P}{P^\theta}$. Removing the terms that are independent of $\theta$ we have
 \begin{widetext}
 \begin{equation}\label{optimality}
   \nabla_\theta \int\!\int \frac{\beta}{h\bar Z} \EXP{-\frac{\beta}{h}|\COPP x' - \COPP\INC(x)|^2} |\COPP x' - \COPP\INCTH(\COPP x)|^2
         \,d\bar x'\,\mu(x)\dx = \nabla_\theta \int |\INCTH(\COPP x) -   \COPP\INC(x)|^2 \mu(x)\,dx = 0\PERIOD
 \end{equation}
 \end{widetext}
 In other words the minimization of $\ENTRATE{P}{P^\theta}$ is equivalent to the minimization
 \begin{equation}\label{minimization}
  \min_{\theta\in\R^n} \int |\COPP a(x) - \bar a(\COPP(x);\theta)|^2\mu(x)\,dx\PERIOD
 \end{equation}
 The minimization becomes particularly straightforward when the parametrization of the coarse-grained drift is 
 chosen as an approximation over the set of polynomials $\{\phi_k(\bar x)\}_{k=1}^n$, i.e., $\bar a(\bar x;\theta) = \sum_k \theta_k \phi_k(\bar x)$.
 In such a case the minimization of the entropy rate functional defines the projection on the subspace $\mathrm{span}\{\phi_k(\bar x)\}_{k=1}^n$ in the
 space $L^2(\mu)$, i.e., the least-square fit with respect to the stationary measure $\mu$. The functional $\ENTRATE{P}{P^\theta}$ is then convex in $\theta$
 and the problem has the unique solution $\mathbf{\theta}^* = (\theta_1^*,\dots,\theta_n^*)$ which is the solution of the linear system
 $$
 \mathbf{\Phi} \mathbf{\theta} = \mathbf{a}\,,\;\;\mbox{where $\mathbf{\Phi}_{ij} = \EXPECT_\mu[\phi_i\phi_j]$ and $\mathbf{a}_i = \EXPECT_\mu[\COPP a\phi_i]$.}
 $$
 The expected values can then be estimated as ergodic averages on a single trajectory realization of the original process $X_t$ as $t\to\infty$.
}
The  parametrization scheme in \cite{Espanol:11} proposed   in the context of  the Fokker-Planck  equation is  an  example of the proposed method 
for reversible stochastic differential equations, i.e., those having a Gibbs steady state. 
{  Application of the proposed method to the diffusion process also shows that widely
applied ``force-matching'' method, \cite{IzvekovVoth:05a,IzvekovVoth:05b,IzvekovVoth:05c},
used in computational coarse-graining is the best-fit in the sense of entropy rate minimization.
}
 }
\end{remark}


%
%
\subsection{Path-space likelihood methods and data-based parametrization of  coarse-grained dynamics}\label{data}

A different, and asymptotically equivalent  perspective on parametrizing coarse-grained dynamics relies on viewing the microscopic simulator 
as means of producing statistical data in the form of
a time-series.
Although the proposed method can be applied to systems simulated by Langevin-type dynamics we demonstrate its application in the  
Kinetic Monte Carlo algorithms in Section~\ref{benchmark}. 
{  The  primary  new element of the presented coarse-graining approach  lies in deriving the parametrization  by  optimizing the information 
content (in  path-space) compared to the available {\em computational data} from a fine-scale simulation, taking advantage of computable formulas 
for relative entropy discussed earlier.}

More specifically, we consider a  fine-scale  data set  of configurations $\mathcal{D}=\{\sigma_1, \sigma_2,...,\sigma_N \}$ obtained for example from a fine-scale 
KMC algorithm.
As is typical in the KMC framework,
we assume that the atomistic model 
can be described by a  spatial, continuous-time  Markov jump process, 
\cite{SV2012}. The  path-space measure of this KMC process, see for the Markov Chain analogue of the path measure \VIZ{path-measure}, 
is parametrized as $P=P^\theta$. In this sense we assume that for the particular data set $\mathcal{D}$ 
the ``true'' parameter value is $\theta=\theta^*$. Identifying   $\theta^*$ amounts, mathematically, to  minimizing 
the pseudo-distance given by the relative entropy,  $\min_\theta \mathcal{R}(P^{\theta^*}|Q^\theta)$. 
Furthermore, following \VIZ{pathrelentropy} it suffices to minimize
$\ENTRATE{P^{\theta^*}}{Q^{\theta}}$. 
On the other hand, using the ergodicity of the fine scale process associated with  the data set $\mathcal{D}=\{\sigma_1, \sigma_2,...,\sigma_N \}$, 
we have the estimators 
\begin{equation}
\label{MLE0}
\ENTRATE{P^{\theta^*}}{Q^{\theta}}=\lim_{N \to \infty} \hat{\mathcal H}_N(P^{\theta^*} |\, Q^{\theta})
\end{equation}
where we define the unbiased estimator for RER, see Section~\ref{stat_est},  
\begin{equation}\label{Path-Lestim}
\hat{\mathcal H}_N(P^{\theta^*} |\, Q^{\theta}):=\frac{1}{N}\sum_{i=1}^N\log\frac{p^{\theta^*}(\sigma_i,\sigma_{i+1})}{q^{\theta}(\sigma_i,\sigma_{i+1})}\, ,
\end{equation}
and $q^{\theta}(\sigma, \sigma')$ is defined in \VIZ{path-measure-CG}.
For simplicity in notation we only demonstrate   the estimator \VIZ{MLE0} for the Markov Chain case, where $p^\theta(\sigma, \sigma')$ denotes the transition probability. 
The continuous-time case, which is relevant to the coarse-grained KMC simulations in Section~\ref{benchmark}, is obtained similarly using \VIZ{entrate2}.

Therefore, the minimization 
of RER becomes
\begin{equation}\label{MLE}
\begin{aligned}
\min_\theta \hat{\mathcal H}_N(P^{\theta^*} |\, Q^{\theta})= & \max_\theta \frac{1}{N}\sum_{i=1}^N\log q^{\theta}(\sigma_i,\sigma_{i+1}) \\
   & - \frac{1}{N}\sum_{i=1}^N\log p^{\theta^*}(\sigma_i,\sigma_{i+1})\COMMA
\end{aligned}
\end{equation}
which does not require {  (a) a priori the knowledge of $\theta^*$,  (b) the microscopic reconstruction defined by $U(\sigma'|\COP\sigma')$ in \VIZ{recon_uniform2} since 
$U(\sigma'|\COP\sigma')$ is independent of $\theta$.
Therefore, we define  the  {\em coarse-grained path space Likelihood} maximization as
\begin{equation}\label{Path-L}
\max_\theta
L(\theta; \{\sigma_i\}_{i=0}^N):=\max_\theta\frac{1}{N}\sum_{i=1}^N\log 
\BARIT{p}^\theta(\COP\sigma_i,\COP\sigma_{i+1})
\PERIOD
\end{equation}
}
Note that if the transition probabilities in \VIZ{Path-L}  are replaced with a stationary measure and $N$ corresponding 
independent samples $\mathcal{D}=\{\sigma_1, \sigma_2,...,\sigma_N \}$,
then \VIZ{Path-L} becomes the classical Maximum Likelihood Principle (MLE).
In this sense \VIZ{Path-L} is a  Maximum Likelihood for the {  coarse-graining of the stationary time series,    $\BARIT{\mathcal{D}}=\{\COP\sigma_1, \COP\sigma_2,...,\COP\sigma_N \}$}
 of the fine-scale  process, and thus includes dynamics information. 
%
Furthermore, due to the stationarity of the time series,  it allows us (if necessary) to obtain the Markovian best-fit from the dynamical simulation and observations on a single, 
long-time realization of the process.

%
%

%


\noindent
{\it Fisher Information Matrix and Confidence Intervals.}
The  Fisher Information Matrix (FIM)  in \VIZ{FIM:MC} is clearly computable as an ergodic average and   can provide  confidence intervals  for the corresponding estimator 
$\hat \theta_N \approx \theta^*$, based on the {\em asymptotic normality} of the MLE estimator $\hat\theta_N$.
Indeed, under additional mild hypotheses on the samples $\mathcal{D}=\{\sigma_1, \sigma_2,...,\sigma_N \}$,   this general procedure guarantees convergence in analogy to the  
central limit theorem, by employing  a martingale formulation, \cite{Crowder}. In a much simpler context, 
when consecutive  pairs $\sigma_i, \sigma_{i+1}$ in the FIM estimator, e.g., \VIZ{FIM:num:approx:MP}, 
are sampled beyond the decorrelation time 
of the time series, the  usual central limit theorem applies and yields the asymptotic normality result
\begin{equation}\label{identify}
   \hat\theta_N \to \theta^{*}\;\mbox{a.s.}\; \mbox{and $N^{1/2} (\hat\theta_N - \theta^*) \WEAKLY N(0,\FISHERR^{-1}(Q^{\theta^*}))$,}
\end{equation}
where  the variance is determined by the Fisher Information Matrix $\FISHERR\big({{Q^{\theta^*}}}\big)$, or asymptotically by $\FISHERR\big({{Q^{\hat\theta_N}}}\big)$.
Thus estimating the FIM $\FISHERR\big({{Q^{\hat\theta_N}}}\big)$ using \VIZ{FIM:MC} provides rigorous error bars on computed optimal parameter values $\theta^*$. 
\section{Statistical estimators for RER and FIM}\label{stat_est}
The Relative Entropy Rate \VIZ{entrate-optim}, as well as the Fisher Information Matrix \VIZ{FIM:MC} are observables of the
stochastic process and can be estimated as ergodic averages. Thus, both observables are computationally tractable since
they depend only on the local transition quantities. We give explicit formulas for the case of the continuous-time Markov chain.

The first estimator for RER is given by 
\begin{equation}
\begin{aligned}
&\widehat{\mathcal H}_1^{(n)}(P\SEP Q^\theta) =  \frac{1}{T} \sum_{i=0}^{n-1} \Delta\tau_i \Big[ \sum_{\sigma'\in E} c(\sigma_i,\sigma')\\
&\times \log \frac{c(\sigma_i,\sigma')}{c^{\theta}(\sigma_i,\sigma')}
- \big(\lambda(\sigma_i) - \lambda^{\theta}(\sigma_i)\big) \Big]\COMMA
\label{RER:num:approx:MP}
\end{aligned}
\end{equation}
where $\Delta\tau_i$ is an exponential random variable with parameter $\lambda(\sigma_i)$
while $T=\sum_i \Delta\tau_i$ is the total simulation time. The sequence $\{\sigma_i\}_{i=0}^n$
is the embedded Markov chain with transition probabilities $p(\sigma_i,\sigma') =
\frac{c(\sigma_i,\sigma')}{\lambda(\sigma_i)}$ at the step $i$ and $c^\theta(\sigma_i,\sigma')$ are the rates
of the parametrized process, e.g., the coarse-grained rates $c^\theta(\COP\sigma_i,\sigma')$. Notice that the weight
$\Delta\tau_i$ which is the waiting time at the state $\sigma_i$ at each step,  is necessary for the
correct estimation of the observable, \cite{Gillespie76}. 
Similarly, the  estimator for the  FIM is
\begin{widetext}
\begin{equation}
\widehat{\bf F}_1^{(n)} = \frac{1}{T} \sum_{i=0}^{n-1} \Delta\tau_i \sum_{\sigma'\in E} 
c^\theta(\sigma_i,\sigma') \nabla_\theta \log c^\theta(\sigma_i,\sigma') \nabla_\theta \log c^\theta(\sigma_i,\sigma')^T \ .
\label{FIM:num:approx:MP}
\end{equation}
\end{widetext}
The computation of the local transition rates $c(\sigma_i,\sigma')$ for
all $\sigma'\in E$ is needed for the simulation of the jump Markov process when Monte Carlo
methods such as stochastic simulation algorithm (SSA), \cite{Gillespie76} is utilized.
Thus, the estimators $\widehat{\mathcal H}_1^{(n)}$ and $\widehat{\bf F}_1^{(n)}$  present only a minor additional computational
cost in the simulation. 

The second numerical estimator
for RER is based on the Girsanov representation of the Radon-Nikodym derivative
and it is given by 
{
\begin{equation}
\begin{aligned}
\widehat{\mathcal H}_2^{(n)}(P\SEP Q^\theta) = &\frac{1}{n} \sum_{i=0}^{n-1}
 \log \frac{c(\sigma_i,\sigma_{i+1})}{c^{\theta}(\sigma_i,\sigma_{i+1})} \\
 & - \frac{1}{T} \sum_{i=0}^{n-1} \Delta\tau_i \big(\lambda(\sigma_i) - \lambda^{\theta}(\sigma_i)\big)\PERIOD
\label{RER:num:approx:MP2}
\end{aligned}
\end{equation}
}
Similarly we can construct an FIM estimator.
The term  in \VIZ{RER:num:approx:MP2} involving logarithms should not be weighted since the
counting measure is approximated with this estimator. 
Unfortunately, the estimator \VIZ{RER:num:approx:MP2} has the same computational cost as
\VIZ{RER:num:approx:MP} due to the need for the computation of the total rate which
is the sum of the local transition rates. Furthermore, in terms of the variance, the latter
estimator has worse performance due to the discarded sum over the states $\sigma'$. For more details we also refer to \cite{PK2013}.

\section{Benchmarks}\label{benchmark}
{\sc Example I: coarse-grained driven Arrhenius diffusion of interacting particles.}
Non-equilibrium  systems at transient or steady state regimes are 
typical in applied science and  engineering, and are the result of coupling between different physicochemical  mechanisms, driven by external couplings or boundary conditions.
For example   reaction-diffusion systems in heteroepitaxial catalytic materials, \cite{SV2012} are typically non-reversible, i.e., their steady state is not a Gibbs distribution. This is
due to  the fact that different mechanisms between species (reaction, diffusion, adsorption, etc.)  do not have a common Hamiltonian 
and thus a common  invariant distribution. Instead, a steady state distribution exists but it is a NESS and is typically not known.  Similarly   separation processes in microporous materials
are typically driven by boundary conditions and/or coupled flows \cite{zeolites_rev},  so the steady state is not necessarily a Gibbs state, but it is again an unknown NESS.

We demonstrate the proposed methodology by selecting a simple but illustrative system in the latter class of non-equilibrium problems. 
We consider  an example of a driven, {\em non-equilibrium}  diffusion process of interacting particles
formulated as a lattice gas model with spin variables $\sigma(x)\in \{0,1\}$, corresponding to occupied or empty lattice sites  $x\in\Lambda_N$; here $\Lambda_N$ denotes a 
uniform one dimensional lattices with $N$ sites.
This is a prototype driven system 
introduced as a model problem  for the influence of microscopic dynamics to macroscopic behavior in driven separations problems in  \cite{VK}.
This model problem is also intimately related to works 
on the structure of non-equilibrium steady states (NESS),
\cite{EPR1,RT1}, 
as well as to  the general formalism of non-equilibrium statistical mechanics, \cite{BLR,JPR}. 

\begin{figure}[ht]
        \centering
        \subfigure[\, Stationary concentration profiles for different cell sizes $q$ with mean-field interactions. The inset depicts errors at different $q$ estimated by $\Hh$]%
                  {\label{fig:CGerrors}\includegraphics[width=0.45\textwidth, height=0.4\textwidth]{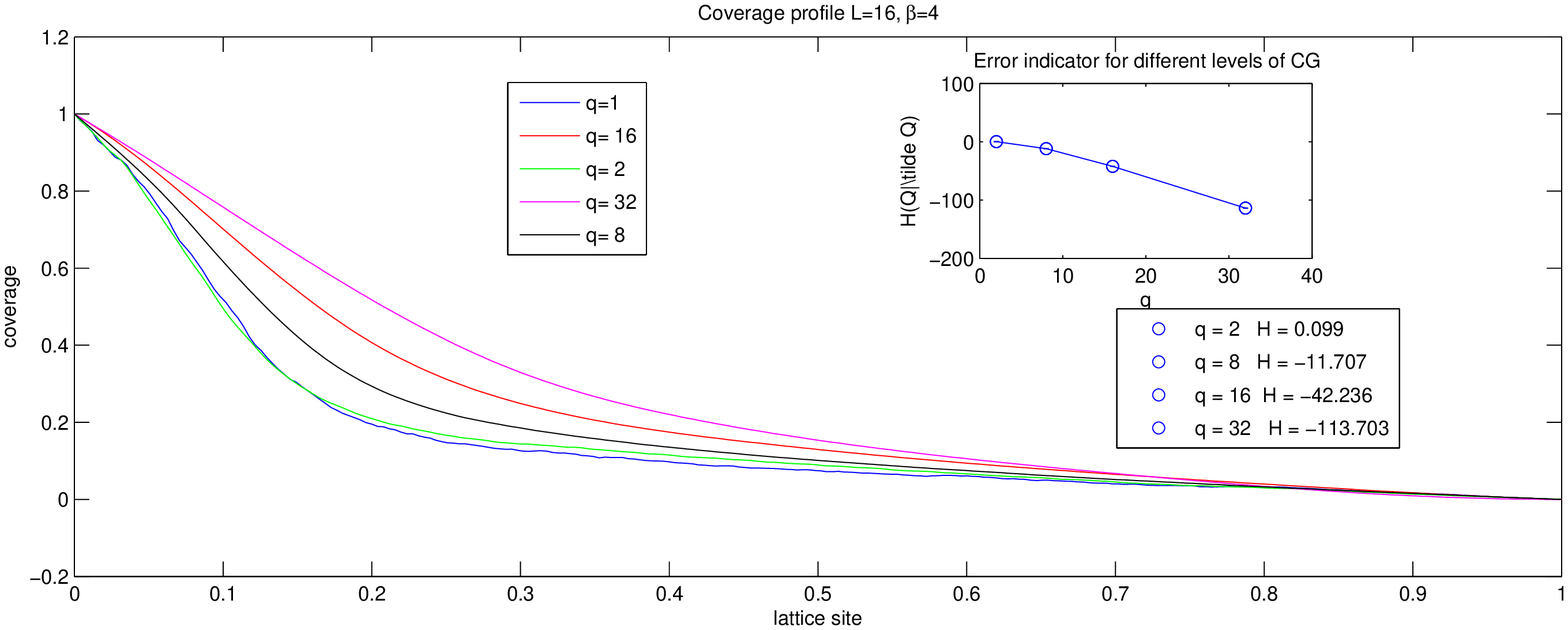}}
          %
        \subfigure[\, Stationary concentration profiles with fitted $\bar J(k;\theta^*)$.]{ \label{fig:fitting}
                \includegraphics[width=0.45\textwidth, height=0.4\textwidth]{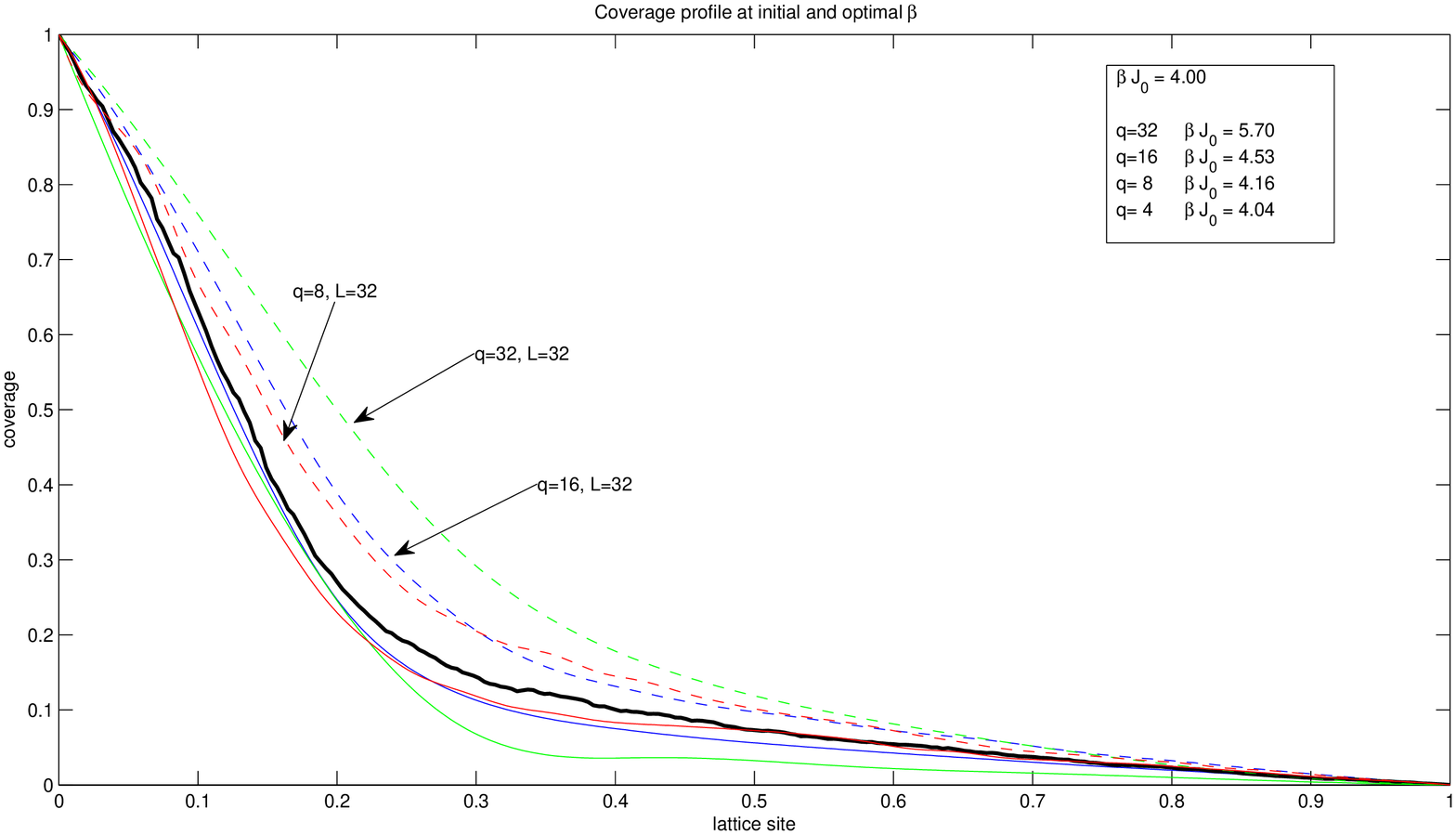}}
\caption{Coarse-grained simulations of driven diffusion of interacting particles without (a) and with (b) fitted effective interactions.}
\label{Fig:NESS}
\end{figure}

The evolution of particles is described in the context of the lattice-gas model as an exchange dynamics with the Arrhenius migration
rate from the site $x\in\LATT$ to the nearest-neighbor sites $|y-x|=1$ 
     $$
          c(x,y,\sigma) = d\, \EXP{-\beta U(x,\sigma)} \sigma(x)(1-\sigma(y))\COMMA
      $$
 which describes the diffusion of a particle at $x$ moving to $y$ 
and  interacting through a two-body potential $J(x-y)$ and  an external field $h$,  defining
an energy barrier $U(x,\sigma) = \sum_{z\neq x} J(x-z)\sigma(z) - h$. 
The continuous-time Markov chain is defined by its rates
and updates to new configurations $\sigma^{x,y}$ in which the spin variables $\sigma(x)$ and $\sigma(y)$ exchanged its values.
Finally, the system is driven by the concentration gradient given by different concentrations at the boundary sites $x=0$ and $x=N$. 
In the long time behavior the distribution converges to a stationary distribution that gives rise to a NESS  concentration
profile across the computational domain\cite{VK}, {  see also Figure~\ref{Fig:NESS}.}

Next we discuss the parametric family of coarse-grained models we will use. Under the assumption of 
{\it a local equilibrium} a straightforward local averaging yields the coarse-grained rates, \cite{KV},
\begin{equation}\label{cgrates}
         \bar c(k,l,\eta) = \frac{1}{q}\eta(k)(q-\eta(l)) d\,\EXP{-\beta \bar U(k,\eta)}\COMMA
\end{equation}
for the lattice-gas model with local concentrations $\eta(k)$ defined as the number of particles
in  a coarse cell $C_k$ of (lattice) size $q$. Thus we have $M$ coarse cells, where  $N=q M$. 
In fact,  according to \VIZ{cgmap} we define  the coarse graining operator 
\begin{equation}\label{cgmap_latt}
\eta(k) = \COP\sigma(k)=\sum_{x\in C_k} \sigma(x)\, ,
\end{equation}
where 
$\eta(k) \in \{0,\dots,q\}$.
Keeping the two-body interactions as a basis for the coarse-grained approximation 
the effective potentials between block spins $k$ and $l$  are obtained by
a straightforward spatial averaging
\begin{equation}\label{barJ}
\begin{aligned}
 \bar J(k,l) &= \frac{1}{q^2} \sum_{x\in C_k}\sum_{y\in C_l} J(x-y)\COMMA \\
 \bar J(k,k) &= \frac{1}{q (q-1)} \sum_{x\in C_k}\sum_{y\in C_k} J(x-y)\PERIOD
\end{aligned}
\end{equation}
Assuming that the approximating dynamics is of Arrhenius type we obtain the energy barriers
      $$
         \bar U(k,\eta) = \sum_l \bar J(k,l) \eta(k) + \bar J(0,0) (\eta(k) - 1) - \bar h\PERIOD
      $$ 
The resulting
dynamics is a Markovian approximation of the coarse-grained evolution and it is defined as CTMC
with the rates $\bar c(k,l,\eta)$. 

As a prototype example of the interactions we consider the constant potential
$J(x) = J_0$ for $|x|\leq L$ and $J(x)=0$ otherwise. This coarse-grained   potential
is parametrized by a single parameter $\theta \equiv \bar J_0$ corresponding to the
strength of the coarse-grained interactions.

First we note that he local mean-field approximation which defines the interaction
potential $\bar J(k-l)$  in \VIZ{barJ} between two block spins $\eta(k)$ and $\eta(l)$ by averaging contributions from all spin-spin interactions
in the cells does not provide a good approximation as demonstrated in Figure~\ref{fig:CGerrors}, where the inset depicts the error estimated
in terms of the entropy rate $\ENTRATE{P}{\tilde P}$. 
However, the mean-field potential $\bar J$ is a good initial datum for \VIZ{iteration}.

\begin{figure}[!htb]
       \centering
        \subfigure[\, Convergence of the estimator $\hat \theta_n$ to the optimal value $\theta^*$.]{\label{fig:convergence}
                  \includegraphics[width=0.4\textwidth]{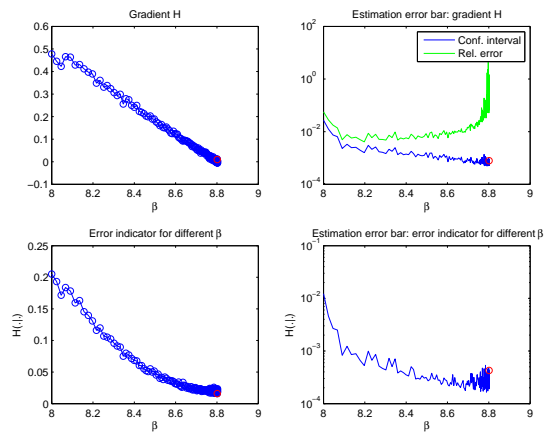}}
        \subfigure[\, Dependence of $\Hh$ on the parameter $\theta$.]{\label{fig:convergenceb}
                  \includegraphics[width=0.4\textwidth]{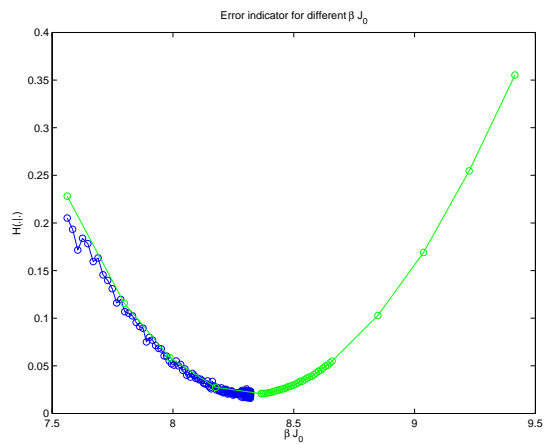}}
\caption{Minimization of $\Hh$. The figure (a) depicts convergence of the estimators for $\hat\theta_n$
         and the gradient (derivative) $\nabla_\theta \Hh$ to the optimal value $\theta^*$ and the optimality
         condition $\nabla_\theta\Hh(\theta^*) = 0$. The right plots depict convergence of 
         confidence intervals for the estimators. The figure (b) demonstrates the convexity of $\Hh$ with
         respect to $\theta$ which holds due to the particular choice $\theta \equiv \beta\bar J_0$ in the
         benchmark.}                
\end{figure}

In this benchmark we stay in the family of two-body potentials and chose to fit only a single parameter that defines
the total strength of the interaction. {  Thus the
rates are parametrized by the effective potential $\bar J(\cdot;\theta)$
using the  single parameter $\theta \equiv \bar J_0$, keeping the interaction range $L$ fixed in each set of simulations. 
Clearly we could consider much richer families with parametrized potentials, e.g., of Morse or Lennard-Jones type, however, we opt for the simplest parametrizations 
in order to demonstrate with clarity  the proposed methods.
The best-fit was obtained by solving  the optimization  problem \VIZ{Path-L}, hence,
minimizing the error defined by $\ENTRATE{P}{\tilde P^\theta}$, see Figure~\ref{fig:convergence} and Figure~\ref{fig:convergenceb}.}
Figure~\ref{fig:fitting} depicts concentration profiles for different sizes $q$ of the coarse cells.
The dashed lines represent results from simulations with mean-field interactions between cells only (i.e., the initial guess in the optimization), 
while the solid lines represent simulations with the parametrized effective interactions. 

Comparison with the profile obtained from the microscopic simulation (the solid black line)
clearly indicates that when the coarse-graining size $q$ becomes close to the interaction range $L$ of the microscopic potential $J$ 
the best-fit in a one-parameter family is {\em not sufficient} for obtaining good approximation and a better candidate class of models, in this case  coarse-grained (CG) dynamics 
$\bar c(k,l,\eta)$, needs to be found
for improved parametrization.
Indeed, in \cite{AKPR, KPRT3} we showed that coarse grained, {\em multi-body  cluster Hamiltonians}  provide such a parametrization.
More specifically, in \cite{KPRT3} we demonstrated, through rigorous cluster expansions that (typical in the state-of-the-art) two-body CG approximations  
break down in lower temperatures and/or for  short range particle-particle 
interactions,  and additional   multi-body CG terms need to be included in the models  in order the CG model  to capture accurately 
phase transitions and other physical important properties. 
Hence, a specific parametrization cannot consistently address this issue on its own, unless the proper  classes of 
parametric models are first  identified. The RER computations such as the one depicted in Figure~\ref{fig:convergenceb} can assess the accuracy of  
different coarse-grained dynamics  within the same 
parametric family (shown here), as well as determine  the comparative coarse-graining accuracy of   different parametric families. 



%
%
%

%
{ 
\medskip
\noindent{\sc Example II: Two-scale diffusion process and averaging principle.}
In the second example we demonstrate the parametric approximation of the coarse-grained process on a system
of two stochastic differential equations with slow and fast time scales
\begin{eqnarray}\label{SDE2scale}
 && dX^\epsilon_t = a(X^\epsilon,Y^\epsilon) dt + dW^1_t \\
 && dY^\epsilon_t = \epsilon^{-1} b(X^\epsilon,Y^\epsilon) dt + \epsilon^{-1/2} dW^2_t\COMMA
\end{eqnarray}
where $W^1_t$, $W^2_t$ are independent standard Wiener processes. Under suitable assumptions on 
$a$, $b$  (see \cite{WeinanCPAM,FreidlinWentzell}) the dynamics $Y^\epsilon_t$ with $x$ held fixed has a unique
invariant measure $\mu^\epsilon_x(dy)$ and as $\epsilon\to 0$ we have 
the effective dynamics for the process $\bar X_t$ given by 
\begin{equation}\label{SDEeff}
 d\bar X_t = \bar a(\bar X_t) + dW_t \PERIOD
\end{equation}
The drift $\bar a$ is given by the averaging principle
$$
   \bar a(x) = \lim_{\epsilon\to 0} \int a(x,y) \,\mu_x^{\epsilon}(dy)\PERIOD
$$
In this example we choose the component $x$ as the coarse variable and compute an approximation of the
drift $\bar a^\epsilon(x)$ for the coarse-grained (projected) process $X^\epsilon_t$. 
The averaging principle suggests that $\bar a^\epsilon \to \bar a$ as $\epsilon\to 0$.

For the specific choice
$$
 a(x,y) = -y\COMMA\;\;\mbox{and}\;\;\;b(x,y) = y - x\COMMA
$$
we have that $\mu_x^\epsilon(dy) = \frac{1}{Z} \exp(-\frac{1}{2}(y-x)^2) dy$ where $Z$ is the normalizing constant and
thus $\bar a(x) = -\frac{1}{Z}\int y \exp(-\frac{1}{2}(y-x)^2) dy = -x$  which yields the effective dynamics in the limit $\epsilon\to 0$.
In the computational benchmarks we thus compare stationary processes  resulting as $t\to\infty$ in solving    
\begin{eqnarray}
 && dX^\epsilon_t = - Y^\epsilon_t dt + dW^1_t\COMMA\;\;\mbox{the two-scale model,} \label{Exorig} \\
 && dY^\epsilon_t = \epsilon^{-1} (Y^\epsilon_t - X^\epsilon_t) dt + \epsilon^{-1/2} dW^2_t\COMMA \nonumber \\
 && d\bar X^\epsilon_t = - \sum_{k=1}^K \theta_k \phi(\bar X^\epsilon_t)\, dt + dW_t\COMMA\;\;\mbox{CG model,} \label{ExAve} \\
 && d\bar X_t = - \bar X_t dt + dW_t\COMMA\;\;\mbox{asymptotic at $\epsilon\to 0$.} \label{ExCGfit} 
\end{eqnarray}
The proposed method for approximating the coarse-grained dynamics constructs an effective potential
$\bar a^\epsilon(x) = \sum_k \theta_k \phi_k(x)$ for a finite value $\epsilon >0$. The set of interpolating
polynomials has been chosen to be $\{1,x,x^2,x^3,x^4\}$ in this example. It is expected that as $\epsilon\to 0$, the approximation
$\bar a^\epsilon$ approaches the averaged coefficient $\bar a(x) = -x$.

The simulation results depicted in Figure~\ref{fig:autocorr1} demonstrate that in the case of sufficient time-scale separation, $\epsilon = 0.005$,
the coarse-grained model well approximates the invariant
distribution as well as the autocorrelation function for stationary dynamics of the component $X_t$. Furthermore, the drift $\bar a^\epsilon(x)$
of the coarse-grained dynamics deviates by a small error (see Figure~\ref{fig:forcefields1}) from the drift $\bar a(x)$ of the averaged model.
On the other hand, for a larger value $\epsilon=0.5$, where the averaging principle does not apply, the proposed method yields a coarse-grained
model (an ``effective dynamics'') which still approximates reasonably well the invariant distribution. However, the dynamics of the stationary
process $\bar X^\epsilon_t$ does not approximate properly the stationary behaviour of the slow component $X^\epsilon_t$ of the original process,
as demonstrated in Figure~\ref{fig:autocorr2} by comparison of autocorrelation functions. 

\begin{figure}[!htb]
       \centering
       \includegraphics[width=0.45\textwidth]{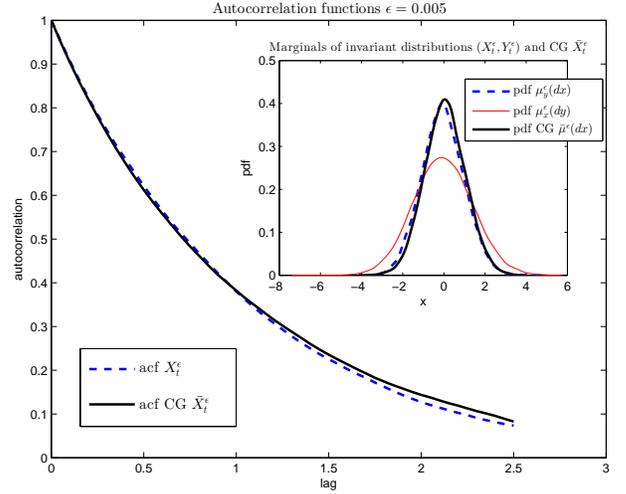}
\caption{Autocorrelation function of the coarse-grained stationary process $\bar X^\epsilon_t$ solving \VIZ{ExCGfit} 
                   for $\epsilon=0.005$ and the autocorrelation function of the process $X^\epsilon_t$ solving \VIZ{Exorig}. The inset depicts the
                   invariant distribution for $\bar X^\epsilon_t$ and marginals of the invariant distribution of $X^\epsilon_t, Y^\epsilon_t$.\label{fig:autocorr1}}                 
\end{figure}

\begin{figure}[!htb]
       \centering
       \includegraphics[width=0.45\textwidth]{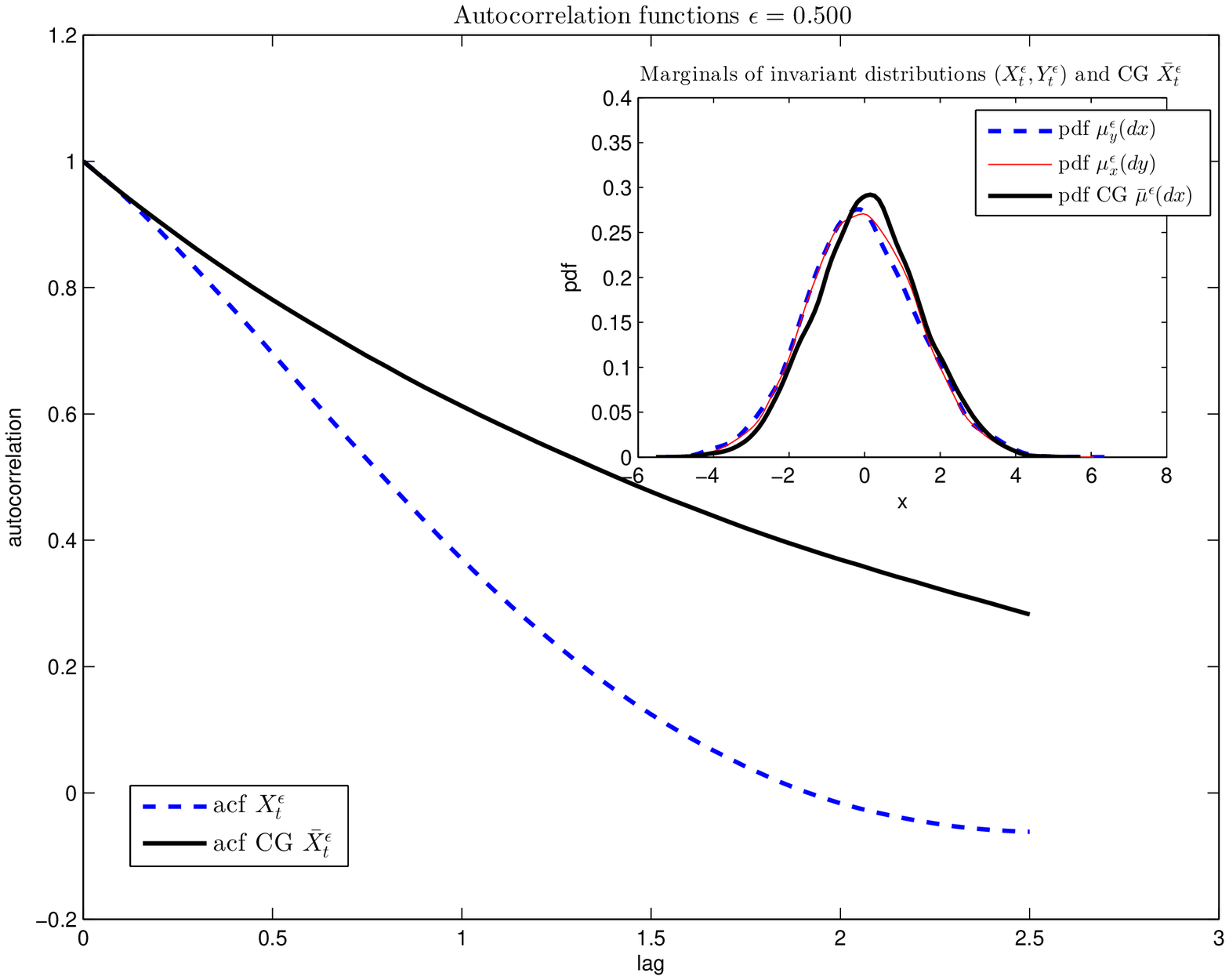}
\caption{Autocorrelation function of the coarse-grained stationary process $\bar X^\epsilon_t$ solving \VIZ{ExCGfit}
                   for $\epsilon=0.5$ and the autocorrelation function of the process $X^\epsilon_t$ solving \VIZ{Exorig}. The inset depicts the
                   invariant distribution for $\bar X^\epsilon_t$ and marginals of the invariant distribution of $X^\epsilon_t, Y^\epsilon_t$.\label{fig:autocorr2}}                
\end{figure}

\begin{figure}[!htb]
       \centering
       \includegraphics[width=0.45\textwidth]{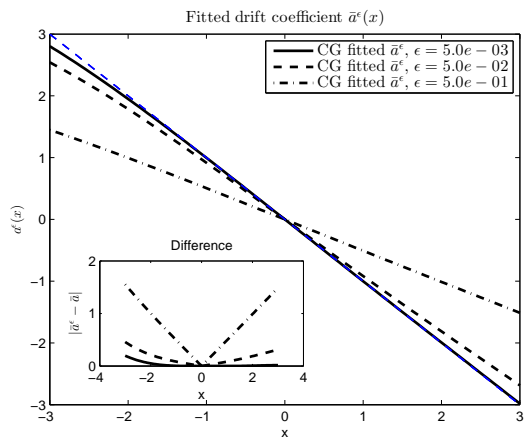}
\caption{Comparison of fitted coefficients $\bar a^\epsilon(x)$ in the coarse-grained model \VIZ{ExCGfit} for different values of $\epsilon$ with the averaged
         drift $\bar a(x)$ in the asymptotic model \VIZ{ExAve}. \label{fig:forcefields1}}                
\end{figure}

\begin{remark}
{\rm
 The fact that the approximating coarse-grained process $\bar X^\epsilon_t$ is asymptotically close to the process $\bar X_t$ resulting from
 the averaging principle is a natural consequence of the proposed fitting method. We give here only a brief heuristic justification: 
 the invariant distribution of the system \VIZ{SDE2scale}
 is $\mu^\epsilon(dx\,dy) = \bar\mu^\epsilon(dx)\mu(dy|x)$ where $\bar\mu^\epsilon$ is the (unknown) stationary distribution of a hidden effective
 dynamics. Under the assumption $\epsilon \ll 1$ we have $\mu^\epsilon(dx\,dy) \approx \bar\mu(dx)\mu_x(dy)$ where $\mu_x(dy)$
 is the invariant distribution of the process $Y^\epsilon_t$ for fixed $x$. Thus the optimization problem \VIZ{minimization} becomes for $\epsilon\to 0$
  \begin{equation}\label{minimization2}
  \min_{\bar a} \int |a(x,y) - \bar a(x)|^2 \mu_x(dy)\bar\mu(x) \,dx\COMMA
 \end{equation}
 which has the unique minimizer
 \begin{equation}
   \bar a(x) = \int a(x,y) \mu_x(dy)\PERIOD
 \end{equation}
 }
\end{remark}

}

\section{Conclusions}\label{concl}
We developed   parametrized model reduction   
methods with controlled-fidelity 
 for  the efficient and reliable coarse-graining  of  {\it non-equilibrium}  molecular systems. 
Such systems  are 
commonplace across all molecular and multi-physics   models  and arise as  the result of coupling between different physical mechanisms, scales, external forcing, and boundary conditions. 
The methodology is  based on concepts from path space information theory  such as the Relative Entropy Rate (RER) and  
allows for constructing optimal parametrized Markovian  coarse-grained dynamics within a given parametric family. The identification of the optimal parameters is achieved  
                  by minimizing information loss--due to coarse-graining--on the path space.       
Furthermore,   a  path-space  analogue of the Fisher Information Matrix   can be derived from RER and provides  confidence intervals  for the corresponding parameter estimators.   
We demonstrate the proposed  coarse-graining methods in 
 (a) non-equilibrium systems with  diffusing  interacting particles, driven by out-of-equilibrium boundary conditions,
and  (b) multi-scale diffusions and  their  corresponding stochastic averaging limits,       
and  show that the proposed RER-based methodology  can assess and improve  the accuracy of  different coarse-grained dynamics  within the same 
parametric family. 
Finally, we expect that  by  employing  systematically derived, e.g., via cluster expansions,  \cite{AKPR, KPRT3},
classes of parametric  families of models,  we can systematically  assess  accuracy vs. computational cost  of  more complex   coarse-grained models, e.g., by  including computationally costly multi-body interactions.  
 The latter issue will be addressed   in  upcoming work.
{ 
\appendix
%
%
\section{Microscopic reconstruction}\label{appdx} 

In this Appendix we discuss the reconstruction procedure  in \VIZ{recon_uniform}. Reversing the coarse-graining, i.e., reproducing microscopic (fine-scale)
properties, directly from coarse-grained (CG)  simulations is an issue that arises
extensively in the coarse-graining  literature, e.g., \cite{mp}.  The
principal idea is that computationally inexpensive CG simulations will
reproduce the large-scale structure and subsequently microscopic
information will be added through {\em microscopic
reconstruction}.  Current approaches address primarily the equilibrium case and rely on
 conditioning on  CG variables and subsequently 
carrying out a local equilibrium relaxation of the microscopic system.

We next  provide a general mathematical framework for the  reconstruction of fine-scale  distributions or transition probabilities  \VIZ{recon_uniform},
from coarse-scale models. For concreteness we focus on reconstructing a fine-scale probability measure 
 ${\mu}_{N}(d\sigma)$ defined on the fine-grained configuration space $\Sigma$, 
 from a CG probability measure $\bar \mu_{M}^{\rm app}(d \eta)$  defined on the coarse space $\bar \Sigma$. 
 
First we define
\[\bar{\mu}_{M}(d\eta)= \int_{\{\sigma:\COP\sigma=\eta \}} {\mu}_{N}(d\sigma)
\]
as the exact coarse-grained measure defined also in \VIZ{exact_CG}.
Then, through the relation
\begin{equation}\label{condition}
{\mu}_{N}(d \sigma)   \equiv\, \mu_N(d\sigma\vert \eta) \bar{\mu}_{M}(d \eta) \, ,
\end{equation}
we  define   the conditional probability $\mu_N(d\sigma\vert\eta)$. In the sense of \VIZ{exact_CG}, we can view  $\mu_N(d\sigma\vert\eta)$  as  the (perfect) reconstruction of  
$\mu_N(d\sigma)$ from  the exactly CG measure $\bar \mu_M(d \eta)$ defined in \VIZ{exact_CG}.
Although many fine-scale configurations 
$\sigma$ correspond to a single CG configuration $\eta$,
the ``reconstructed" conditional probability measure $\mu_N(d\sigma\vert\eta)$ is {\em uniquely} defined, 
given the microscopic and the coarse-grained measures $\mu_N(d\sigma)$ and  $\bar \mu_M(d \eta)$ respectively.

A coarse-graining scheme provides an approximation $\bar \mu_{M}^{\rm app}(d \eta)$ 
for $\bar \mu_M(d \eta)$.  The approximation  $\bar \mu_{M}^{\rm app}(d\eta)$ could be, for instance,    the schemes discussed in Section \ref{benchmark}. 
To provide a reconstruction we need to lift the measure $\bar \mu_{M}^{\rm app}(d \eta)$ to
a measure $\mu_{N}^{\rm app}(d \sigma)$ on the microscopic configurations. That is,  we  
need to specify a conditional probability $\nu_N(d\sigma \vert  \eta)$ and set
\begin{equation}\label{recon}
\mu_N^{\rm app}(d\sigma)\,:=\, \nu_N(d\sigma \vert  \eta) \bar \mu_{M}^{\rm app}(d\eta)\,.
\end{equation}
In the spirit of our earlier discussion on using relative entropy to quantify  the quality of approximation in CG schemes, 
it is natural to measure the efficiency of the reconstruction  by the specific relative entropy
${\cal R} \left( \mu_N^{\rm app}  \vert {\mu}_{N} \right)$. A simple computation\cite{Cover} shows
that
\begin{widetext}
\begin{equation}\label{entdec}
{\cal R} \left( \mu_N^{\rm app}  \SEP {\mu}_{N} \right)\,=\, {\cal R} \left( \bar \mu_M^{\rm app}  \SEP \bar {\mu}_{M} \right) + 
\int  {\cal R} \left( \nu_N( \cdot \vert \eta) \SEP \mu_N( \cdot \SEP \eta) \right)  \bar{\mu}_{M}^{\rm app}(d \eta) 
\,,
\end{equation}    
\end{widetext} 
i.e., relative entropy splits the total error at the microscopic level into the sum of the error at the coarse level and the 
error made during the reconstruction.  

The first term in \eqref{entdec} can be controlled, for example, by error analysis results on the coarse variables\cite{KPRT1, KPRT3}.   In order to obtain a suitable  reconstruction we then need to construct $\nu_N(d\sigma\,\vert\,\eta)$ such that (a) it is easily computable and implementable, and 
(b) the error ${\cal R} \left( \nu_N( d\sigma\,\vert\, \eta) \SEP \mu_N( d\sigma\SEP \eta) \right)$ 
should be of the same order as the first term in  \eqref{entdec}. 

\medskip
\noindent
{\sc Example:} 
The simplest example of reconstruction for a microscopic system $\mu(d\sigma)$ with a coarse-grained probability distribution $\bar{\mu}_M^{\rm app}(d \eta)$ is obtained by
\begin{equation}\label{recon0a}
\mu^{\rm app}(\sigma)\,:=\, U(\sigma \vert  \eta) \bar{\mu}^{\rm app}(\eta)\, , 
\end{equation}
 where $U(\sigma'|\eta')={1 \over |\{\sigma:\COP\sigma=\eta'\}|}$ is the uniform conditional distribution over all fine-scale states $\sigma$ corresponding to the same coarse-grained state $\eta'$.
i.e., we first sample the CG variables $\eta$  using the CG probability; then  we reconstruct the microscopic configuration $\sigma$ by distributing the particles uniformly on the 
coarse cell, conditioned on the value of $\eta$.  More accurate, but computationally more demanding schemes were proposed in \cite{TT, KKPV}.
}
\section*{Acknowledgements}

The research of M.A.K. was supported by the National Science Foundation through the  CDI -Type II award NSF-CMMI-0835673
and  by the European Union (European Social Fund) and Greece (National Strategic Reference Framework), under the THALES Program, grant
AMOSICSS. The research of P.P. was partially supported by the National Science Foundation under the CDI -Type II award
NSF-CMMI-0835582. { M.A.K. also acknowledges numerous discussions with Matthew Dobson, Yannis Pantazis and Luc Rey-Bellet.
The authors would like to thank the anonymous referees for their valuable comments.}


%

\end{document}